# A Polynomial kernel for Proper Interval Vertex Deletion

Fedor V. Fomin[1], Saket Saurabh[2], and Yngve Villanger[1]

[1] Department of Informatics, University of Bergen, N-5020 Bergen, Norway.
{fomin|yngvev}@ii.uib.no
[2] The Institute of Mathematical Sciences, Chennai, India
saket@imsc.res.in

**Abstract.** It is known that the problem of deleting at most $k$ vertices to obtain a proper interval graph (Proper Interval Vertex Deletion) is fixed parameter tractable. However, whether the problem admits a polynomial kernel or not was open. Here, we answers this question in affirmative by obtaining a polynomial kernel for Proper Interval Vertex Deletion. This resolves an open question of van Bevern, Komusiewicz, Moser, and Niedermeier.

## 1 Introduction

Study of graph editing problems cover a large part of parmeterized complexity. The problem of editing (adding/deleting vertices/edges) to ensure a graph to have some property is a well studied problem in theory and applications of graph algorithms. When we want the *edited graph* to be in a hereditary (that is, closed under induced subgraphs) graph class, the optimization version of the corresponding node/edge deletion problems are known to be $NP$-complete by a classical result of Lewis and Yannakakis [17]. In this paper we study the problem of deleting vertices to get into proper interval graph in the realm of kernelization complexity.

A graph $G$ is a proper (unit) interval graph if it is an intersection graph of unit-length intervals on a real line. Proper interval graphs form a well studied and well structured hereditary class of graphs. The parameterized study of the following problem of deleting vertices to get into proper interval graph was initiated by van Bevern et al. [23].

---

$p$-Proper Interval Vertex Deletion (PIVD)  **Parameter:** $k$
**Input:** An undirected graph $G$ and a positive integer $k$
**Question:** Decide whether $G$ has a vertex set $X$ of size at most $k$ such that $G \setminus X$ is a proper interval graph

---

Wegner [25] (see also [3]) showed that proper interval graphs are exactly the class of graphs that are $\{claw, net, tent, hole\}$-free. *Claw, net*, and *tent* are graphs containing at most 6 vertices depicted in Fig. 1, and *hole* is an induced cycle of length at least four. Combining results of Wegner, Cai, and Marx [4,19,25], it can be shown that PIVD is FPT. That is one can obtain an algorithm for PIVD running in time $\tau(k)n^{O(1)}$ where $\tau$ is a function depending only on $k$ and $n$ is the number of vertices in the input graph. Van Bevern et al. [23] presented a faster $O(k(14k+14)^{k+1}n^6)$ time algorithm for PIVD using the structure of a problem instance that is already $\{claw, net, tent, C_4, C_5, C_6\}$-free. The running time was recently improved by Villanger down to $O(6^k k n^6)$ [24]. However, the question, whether the problem has a polynomial kernel or not was not resolved. This question was explicitly asked by Van Bevern et al. [23]. This is precisely the problem we address in this paper.

Here, we study PIVD from kernelization perspective. A parameterized problem is said to admit a *polynomial kernel* if every instance $(I, k)$ can be reduced in polynomial time to an equivalent instance with both size and parameter value bounded by a polynomial in $k$. In other words, it is possible in polynomial time to "compress" every instance of the problem to a new instance of size $k^{\mathcal{O}(1)}$.

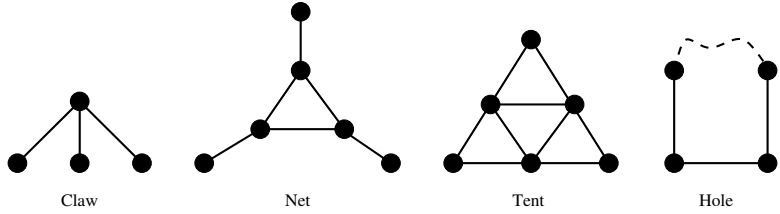

**Fig. 1.** Excluded Subgraphs

The study of kernelization is one of the leading research frontier of modern parameterized complexity and the major recent advances in the area are on kernelization. Important recent developments are the introduction of new lower bounds techniques, showing (under complexity theoretic assumptions) that certain problems must have kernels of at least certain sizes [2,5,10] and general results showing that large classes of problems have small (e.g., linear) kernels [1,9,15]. Identifying the intractability borders inside the class FPT is an important challenge in parameterized complexity and kernelization provides a way to do it as we can classify the problems based on what sizes kernel they admit. From practical side, kernelization algorithms often lead to efficient preprocessing rules which can significantly reduce and simplify the initial instance [14,26]. Thus, kernelization provides a framework for the mathematical analysis of polynomial time preprocessing.

Our interest to PIVD is also motivated by the following more general problem. Let $\mathcal{G}$ be an arbitrary class of graphs. We denote by $\mathcal{G} + kv$ the class of graphs that can be obtained from a member of $\mathcal{G}$ by adding at most $k$ vertices. For an example, PIVD is equivalent to deciding if $G$ is in $\mathcal{G} + kv$, where $\mathcal{G}$ is the class of proper interval graphs. There is a generic criteria providing sufficient conditions on the properties of class $\mathcal{G}$ to admit a polynomial kernel for $\mathcal{G} + kv$ recognition problem. A graph class is called *hereditary* if every induced subgraph of every graph in the class also belongs to the class. Let $\Pi$ be a hereditary graph class characterized by forbidden induced subgraphs of size at most $d$. Cai [4] showed that the $\Pi + kv$ problem, where given an input graph $G$ and positive integers $k$, the question is to decide whether there exists a $k$ sized vertex subset $S$ such that $G[V \setminus S] \in \Pi$, is FPT when parameterized by $k$. The $\Pi + kv$ problem can be shown to be equivalent to $p$-$d$-HITTING SET and thus it admits a polynomial kernel [16]. In the $p$-$d$-HITTING SET problem, we are given a family $\mathcal{F}$ of sets of size at most $d$ over a universe $\mathcal{U}$ and a positive integer $k$ and the objective is to find a subset $S \subseteq \mathcal{U}$ of size at most $k$ intersecting, or *hitting*, every set of $\mathcal{F}$.

However, the result of Cai does not settle the parameterized complexity of $\Pi + kv$ when $\Pi$ cannot be characterized by a finite number of forbidden induced subgraphs. Here even for graph classes with well-understood structure and very simple infinite set of forbidden subgraphs, the situation becomes challenging. In particular, for the "closest relatives" of proper interval graphs, chordal and interval graphs, the current situation is still obscure. For example, the FPT algorithm of Marx [19] for the problem of vertex deletion into a chordal graph, i.e. a graph without induced cycles of length at least four, requires heavy algorithmic machinery. The question if CHORDAL+$kv$ admits a polynomial kernel is still open. Situation with INTERVAL+$kv$ is even more frustrating, in this case we even do not know if the problem is in FPT or not.

In this paper we make a step towards understanding the kernelization behaviour for $\mathcal{G}+kv$ recognition problems, where $\mathcal{G}$ is well understood and the infinite set of forbidden subgraphs is simple. A generic strategy to obtain FPT algorithm for many $\mathcal{G} + kv$ recognition problems is to first take care of small forbidden subgraphs by branching on them. When these small subgraphs are not present, the structure of a graph is utilised to take care of infinite family of forbidden subgraphs. However, to apply a similar strategy for kernelization algorithm we need to obtain a polynomial kernel for a variant of $p$-$d$-HITTING SET that preserves all minimal solutions of size at most $k$ along with a witness for the minimality, rather than the kernel for $p$-$d$-HITTING SET which was sufficient when we only had finitely many forbidden induced subgraphs. Preserving the witness for the minimality is crucial here as this "insulate" small constant size forbidden induced subgraphs from the large and infinite forbidden induced subgraph. In some way it mimics the generic strategy used for the FPT algorithm. Towards this we show that indeed one can obtain a kernel for variant of $p$-$d$-HITTING SET that preserves all minimal solutions of size at most $k$ along with a witness for the minimality (Section 3). Finally, using this in combination with reduction rules that shrinks "clique and clique paths" in proper interval graphs we resolve the kernelization complexity of PIVD. We show that PIVD admits a polynomial kernel and thus resolving the open question posed in [23]. We believe that our strategy to obtain polynomial kernel for PIVD will be useful in obtaining polynomial kernels for various other $\mathcal{G} + kv$ recognition problems.

## 2 Definitions and notations

We consider simple, finite, and undirected graphs. For a graph $G$, $V(G)$ is the *vertex set* of $G$ and $E(G)$ is the *edge set* of $G$. For every edge $uv \in E(G)$, vertices $u$ and $v$ are *adjacent* or *neighbours*. The *neighbourhood* of a vertex $u$ in $G$ is $N_G(u) = \{v \mid uv \in E\}$, and the *closed neighbourhood* of $u$ is $N_G[u] = N_G(u) \cup \{u\}$. When the context will be clear we will omit the subscript. A set $X \subseteq V$ is called *clique* of $G$ if the vertices in $X$ are pairwise adjacent. A *maximal* clique is a clique that is not a proper subset of any other clique. For $U \subseteq V$, the *subgraph of $G$ induced by $U$* is denoted by $G[U]$ and it is the graph with vertex set $U$ and edge set equal to the set of edges $uv \in E$ with $u, v \in U$. For every $U \subseteq V$, $G' = G[U]$ is an *induced subgraph* of $G$. By $G \setminus X$ for $X \subseteq V$, we denote the graph $G[V \setminus X]$.

**Parameterized problems and kernels.** A parameterized problem $\Pi$ is a subset of $\Gamma^* \times \mathbb{N}$ for some finite alphabet $\Gamma$. An instance of a parameterized problem consists of $(x, k)$, where $k$ is called the parameter. The notion of kernelization is formally defined as follows. A *kernelization algorithm*, or in short, a *kernelization*, for a parameterized problem $\Pi \subseteq \Gamma^* \times \mathbb{N}$ is an algorithm that, given $(x, k) \in \Gamma^* \times \mathbb{N}$, outputs in time polynomial in $|x| + k$ a pair $(x', k') \in \Gamma^* \times \mathbb{N}$ such that (a) $(x, k) \in \Pi$ if and only if $(x', k') \in \Pi$ and (b) $|x'|, k' \leq g(k)$, where $g$ is some computable function depending only on $k$. The output of kernelization $(x', k')$ is referred to as the *kernel* and the function $g$ is referred to as the size of the kernel. If $g(k) \in k^{O(1)}$, then we say that $\Pi$ admits a polynomial kernel. For general background on the theory, the reader is referred to the monographs [6,8,20].

**Interval graphs.** A graph $G$ is an *interval graph* if and only if we can associate with each vertex $v \in V(G)$ an open interval $I_v = (l_v, r_v)$ on the real line, such that for all $v, w \in V(G)$, $v \neq w$: $vw \in E(G)$ if and only if $I_v \cap I_w \neq \emptyset$. The set of intervals $\mathcal{I} = \{I_v\}_{v \in V}$ is called an (interval) *representation* of $G$. By the classical result of Gilmore and Hoffman [12], and Fulkerson and Gross [11], for every interval graph $G$ there is a linear ordering of its maximal cliques such that for every vertex $v$, the maximal cliques containing $v$ occur consequently. We refer to such an ordering of maximal cliques $C_1, C_2, \ldots, C_p$ of interval graph $G$ as a *clique path* of $G$. Note that an interval graph can have several different clique paths. A clique path of an interval graph can be constructed in linear time [11].

A *proper interval graph* is an interval graph with an interval model where no interval is properly contained in any other interval. There are several equivalent definitions of proper interval graphs. Graph $G$ is a *unit interval graph* if $G$ is an interval graph with an interval model of unit-length intervals. By the result of Roberts [22], $G$ is a unit interval graph if and only if it is a proper interval graph. A *claw* is a graph that is isomorphic to $K_{1,3}$, see Fig. 1. A graph is *claw-free* if it does not have a claw as an induced subgraph. Proper interval graphs are exactly the claw-free interval graphs [22].

A vertex ordering $\sigma = \langle u_1, \ldots, u_n \rangle$ of graph $G = (V, E)$ is called *interval ordering* if for every $1 \leq i < j < k \leq n$, $v_i v_k \in E$ implies $v_j v_k \in E$. A graph is an interval graph if and only if it admits an interval ordering [21]. A vertex ordering $\sigma$ for $G$ is called a *proper interval ordering* if for every for every $1 \leq i < j < k \leq n$, $v_i v_k \in E$ implies $v_i v_j, v_j v_k \in E$. A graph is a proper interval graph if and only if it admits a proper interval ordering [18]. Interval orderings and proper interval orderings can be computed in linear time, if they exist. We will need the following properties of proper interval graphs.

**Proposition 1 ([25,3]).** *A graph $G$ is a proper interval graph if and only it contains neither claw, net, tent, nor cycles (holes) of length at least 4 as induced subgraphs.*

A *circular-arc graph* is the intersection graph of a set of arcs on the circle. A circular-arc graph is a *proper circular-arc graph* if no arc is properly contained in any other arc.

**Proposition 2 ([24]).** *Every connected graph $G$ that does not contain either tent, net or claw or induced cycles (holes) of length 4, 5 and 6 as an induced subgraph is a proper circular-arc graph. Moreover, there is a polynomial time algorithm computing a set $X$ of minimum size such that $G \setminus X$ is a proper interval graph.*

The following proposition of proper interval orderings of proper interval graphs follows almost directly from the definition.

**Proposition 3.** *Let $\sigma = \langle v_1, \ldots, v_n \rangle$ be a proper interval ordering of $G = (V, E)$.*

1. *For every maximal clique $K$ of $G$, there exist integers $1 \leq i < j \leq n$ such that $K = \{v_i, v_{i+1}, \ldots, v_{j-1}, v_j\}$. That is, vertices of $K$ occur consecutively.*
2. *For a vertex $v_\ell$ let $i, j$ be the smallest and the largest numbers such that $v_i v_\ell, v_\ell v_j \in E$, then $N[v_\ell] = \{v_i, \ldots, v_j\}$ and the sets $\{v_i, \ldots, v_\ell\}$ and $\{v_\ell, \ldots, v_j\}$ are cliques;*
3. *Let $C_1, C_2, \ldots, C_p$ be a clique path of $G$. If $v_i \in C_j$ then $v_i \notin C_{j+\ell+1}$, where $\ell \geq |N[v_i]|$.*

## 3 Sunflower Lemma and minimal hitting sets

In this section we obtain a kernel for a variant of $p$-$d$-HITTING SET that preserves all minimal solutions of size at most $k$ along with a witness for the minimality. Towards this we introduce the notion of *sunflower*. A *sunflower* $S$ with $k$ *petals* and a *core* $Y$ is a collection of sets $\{S_1, S_2, \ldots, S_k\}$ such that $S_i \cap S_j = Y$ for all $i \neq j$; the sets $S_i \setminus Y$ are petals and we require that none of them be empty. Note that a family of pairwise disjoint sets is a sunflower (with an empty core). We need the following algorithmic version of the classical result of Erdős and Rado [7].

**Lemma 1 ([8]). [Sunflower Lemma]** *Let $\mathcal{F}$ be a family of sets over a universe $\mathcal{U}$ each of cardinality at most $d$. If $|\mathcal{F}| > d!(k-1)^d$ then $\mathcal{F}$ contains a sunflower with $k$ petals and such a sunflower can be found in $O(k + |\mathcal{F}|)$ time.*

A subset $X$ of $\mathcal{U}$ intersecting every set in $\mathcal{F}$ is referred to as a *hitting set* for $\mathcal{F}$. Sunflower Lemma is a common tool used in parameterized complexity to obtain a polynomial kernel for $p$-$d$-HITTING SET [8]. The observation is that if $\mathcal{F}$ contains a sunflower $S = \{S_1, \ldots, S_{k+1}\}$ of cardinality $k+1$ then every hitting set of of size at most $k$ of $\mathcal{F}$ must have a nonempty intersection with the core $Y$. However, for our purposes it is crucial that kernelization algorithm preserves *all* small minimal hitting sets. The following application of Sunflower Lemma is very similar to its use for kernelization for $p$-$d$-HITTING SET. However, it does not seem to exist in the literature in the form required for our kernelization and thus we give its proof here.

**Lemma 2.** *Let $\mathcal{F}$ be a family of sets of cardinality at most d over a universe $\mathcal{U}$ and k be a positive integer. Then there is an $\mathcal{O}(|\mathcal{F}|(k+|\mathcal{F}|))$ time algorithm that finds a non-empty set $F' \subseteq \mathcal{F}$ such that*

1. *For every $Z \subseteq \mathcal{U}$ of size at most $k$, $Z$ is a minimal hitting set of $\mathcal{F}$ if and only if $Z$ is a minimal hitting set of $\mathcal{F}'$; and*
2. *$|\mathcal{F}'| \leq d!(k+1)^d$.*

*Proof.* The algorithm iteratively construct sets $\mathcal{F}_t$, where $0 \leq t \leq |\mathcal{F}|$. We start with $t = 0$ and $\mathcal{F}_0 = \mathcal{F}$. For $t \geq 1$, we use Lemma 1 to check if there is a sunflower of cardinality $k+2$ in $\mathcal{F}_{t-1}$. If there is no such sunflower, we stop, and output $\mathcal{F}' = \mathcal{F}_{t-1}$. Otherwise, we use Lemma 1 to construct a sunflower $\{S_1, S_2, \ldots, S_{k+2}\}$ in $\mathcal{F}_{t-1}$. We put $\mathcal{F}_t = \mathcal{F}_{t-1} \setminus \{S_{k+2}\}$. At every step, we delete one subset of $\mathcal{F}$. Thus the algorithm calls the algorithm from Lemma 1 at most $|\mathcal{F}|$ times and hence its running time is $\mathcal{O}(|\mathcal{F}|(k+|\mathcal{F}|))$. Since $\mathcal{F}'$ has no sunflower of cardinality $k+2$, by Lemma 1, $|\mathcal{F}'| \leq d!(k+1)^d$.

Now we prove that for each $t \geq 1$ and for every set $Z \subseteq \mathcal{U}$, it holds that $Z$ is a minimal hitting set for $\mathcal{F}_{t-1}$ of size $k$ if and only if $Z$ is a minimal hitting set for $\mathcal{F}_t$. Since for $t = 1$, $\mathcal{F}_{t-1} = \mathcal{F}$, and for some $t \leq |\mathcal{F}|$, $\mathcal{F}_t = \mathcal{F}'$, by transitivity this is sufficient for proving the first statement of the lemma.

The set $\mathcal{F}_t$ is obtained from $\mathcal{F}_{t-1}$ by removing the set $S_{k+2}$ of the sunflower $\{S_1, S_2, \ldots, S_{k+2}\}$ in $\mathcal{F}_{t-1}$. Let $Y$ be the core of this sunflower. If $Y = \emptyset$, then $\mathcal{F}_{t-1}$ has no hitting set of size $k$. In this case, $\mathcal{F}_t$ contains pairwise disjoint sets $S_1, S_2, \ldots, S_{k+1}$ and hence $\mathcal{F}_t$ also has no hitting set of size $k$. Thus the interesting case is when $Y \neq \emptyset$.

Let $Z$ be a minimal hitting set for $\mathcal{F}_{t-1}$ of size $k$. Since $\mathcal{F}_t \subseteq \mathcal{F}_{t-1}$, we have that set $Z$ is a hitting set for $\mathcal{F}_t$. We claim that $Z$ is a *minimal* hitting set for $\mathcal{F}_t$. Targeting towards a contradiction, let us assume that $Z$ is not a minimal hitting set for $\mathcal{F}_t$. Then there is $u \in Z$, such that $Z' = Z \setminus \{u\}$ is a hitting set for $\mathcal{F}_t$. Sets $S_1, S_2, \ldots, S_{k+1}$ form sunflower in $\mathcal{F}_t$, and thus every hitting set of size at most $k$, including $Z'$, intersects its core $Y$. Thus $Z'$ hits all sets of $\mathcal{F}_{t-1}$, as it hits all the sets of $\mathcal{F}_t$ and it also hits $S_{k+2}$ because $Y \subset S_{k+2}$. Therefore, $Z$ is not a minimal hitting set in $\mathcal{F}_{t-1}$, which is a contradiction. This shows that $Z$ is a minimal hitting set for $\mathcal{F}_t$.

Let $Z$ be a minimal hitting set for $\mathcal{F}_t$ of size $k$. Every hitting set of size $k$ for $\mathcal{F}_t$ should contain at least one vertex of the core $Y$. Hence $Y \cap Z \neq \emptyset$. But then $Z \cap S_{k+2} \neq \emptyset$ and thus $Z$ is a hitting set for $\mathcal{F}_{t-1}$. Because $\mathcal{F}_t \subseteq \mathcal{F}_{t-1}$, $Z$ is a minimal hitting set for $\mathcal{F}_{t-1}$. □

Given a family $\mathcal{F}$ of sets over a universe $\mathcal{U}$ and a subset $T \subseteq \mathcal{U}$, we define $\mathcal{F}_T$ as the subset of $\mathcal{F}$, containing all sets $Q \in \mathcal{F}$ such that $Q \subseteq T$.

## 4 Proper Interval Vertex Deletion

In this section, we apply results from the previous section to obtain a polynomial kernel for PIVD. Let $(G, k)$ be an instance to PIVD, where $G$ is a graph on $n$ vertices and $k$ is a positive integer. The kernel algorithm is given in four steps. First we take care of small forbidden sets using Lemma 2, the second and the third steps reduce the size of maximal cliques and shrink the length of induced paths. Finally, we combine three previous steps into a kernelization algorithm.

### 4.1 Small induced forbidden sets

In this section we show how we could use Lemma 2 to identify a vertex subset of $V(G)$, which allows us to forget about small induced subgraphs in $G$ and to concentrate on long induced cycles in the kernelization algorithm for PIVD. We view vertex subsets of $G$ inducing forbidden subgraph as a set family. We prove the following lemma.

**Lemma 3.** *Let $(G, k)$ be an instance to* PIVD*. Then there is a polynomial time algorithm that either finds a non-empty set $T \subseteq V(G)$ such that*

1. *$G \setminus T$ is a proper interval graph;*
2. *Every set $Y \subseteq V(G)$ of size at most $k$ is a minimal hitting set for nets, tents, claws and induced cycles $C_\ell$, $4 \leq \ell \leq 8$, in $G$ if and only if it is a minimal hitting set for nets, tents, claws and induced cycles $C_\ell$, $4 \leq \ell \leq 8$, contained in $G[T]$; and*
3. *$|T| \leq 8 \cdot 8!(k+1)^8 + k$.*

*or concludes that $(G, k)$ is a* `NO` *instance.*

*Proof.* Let $\mathcal{F}$ be the family consisting of all nets, tents, claws and induced cycles $C_\ell$, $4 \leq \ell \leq 8$, of the input graph $G$. We apply Lemma 2 on $\mathcal{F}$ and in polynomial time find $\mathcal{F}'$ such that

1. $Y$ is a minimal hitting set of $\mathcal{F}$ of size at most $k$ if and only if $Y$ is a minimal hitting set of $\mathcal{F}'$ of size at most $k$; and
2. $|\mathcal{F}'| \leq 8!(k+1)^8 + k$.

We take $T$ to be the elements contained inside any set of $\mathcal{F}'$. Thus $|T| \leq 8 \cdot 8!(k+1)^8$. In graph theoretic terms, this means that every vertex set $Y \subseteq V(G)$ of size at most $k$ is a minimal hitting set for nets, tents, claws and induced cycles $C_\ell$, $4 \leq \ell \leq 8$, contained in $G$ if and only if it is a minimal hitting set for nets, tents, claws and induced cycles $C_\ell$, $4 \leq \ell \leq 8$, contained in $G[T]$. Then $G \setminus T$ contains neither tent, net, claw nor induced cycles of length 4, 5 or 6 and by Proposition 2, is a proper circular-arc graph. Using Proposition 2, in polynomial time we find a minimum size set $X$ of $V(G) \setminus T$ such that $G \setminus (T \cup X)$ is a proper interval graph. If the size of $|X| > k$, then we conclude that $(G, k)$ is a `NO` instance. So we assume that $|X| \leq k$. Now we add $X$ to $T$, increasing its size by at most $k$. This concludes the proof. □

In the rest of the coming subsections we assume that

$$\boxed{G_T = G \setminus T \text{ is a proper interval graph and } |T| \leq \delta(k) = 8 \cdot 8!(k+1)^8 + k.}$$

### 4.2 Finding irrelevant vertices in $G_T$

In this subsection we show that if the maximum size of a clique in $G_T$ is larger than $(k+1)(\delta(k)+2)$, then we can find some irrelevant vertex $v \in V(G_T)$ and delete it without altering the answer to the problem. More precisely, we prove the following result.

**Lemma 4.** *Let $G$ and $T$ be as described before. Furthermore, let the size of a maximum clique in $G_T$ be greater than $\epsilon(k) = (k+1)(\delta(k)+2)$. Then in polynomial time we can find a vertex $v \in V(G_T)$ such that $(G, k)$ is a* `YES` *instance to* PIVD *if and only if $(G \setminus v, k)$ is a* `YES` *instance.*

*Proof.* We start by giving a procedure to find an irrelevant vertex $v$. Let $K$ be a maximum clique of $G_T$, it is well known that a maximum clique can be found in linear time in proper interval graphs [13]. Let $\sigma = \langle u_1, \ldots, u_n \rangle$ be a linear vertex ordering of $G_T$. By Proposition 3, vertices of $K$ form an interval in $\sigma$, we denote this interval by $\sigma(K)$. Suppose that $|K| > \epsilon(k)$. The following procedure marks vertices in the clique $K$ and helps to identify an irrelevant vertex.

Set $Z = \emptyset$. For every vertex $v \in T$, pick $k+1$ arbitrary neighbours of $v$ in $K$, say $S_v$, and add them to $Z$. If $v$ has at most $k$ neighbors in $K$, then add all of them to $Z$. Furthermore, add $V^F$, the first $k+1$ vertices, and $V^L$, the last $k+1$ vertices in $\sigma(K)$ to $Z$. Return $Z$.

Observe that the above procedure runs in polynomial time and adds at most $k+1$ vertices for any vertex in $T$. In addition, the procedure also adds some other $2(k+1)$ vertices to $Z$. Thus the size of the set $Z$ containing marked vertices is at most $(k+1)(\delta(k)+2) = \epsilon(k)$. By our assumption on the size of the clique we have that $K \setminus Z \neq \emptyset$. We show that any vertex in $K \setminus Z$ is irrelevant. Let $v \in (K \setminus Z)$. Now we show that $(G, k)$ is a YES instance to PIVD if and only if $(G \setminus v, k)$ is a YES instance to PIVD. Towards this goal we first prove the following auxiliary claim.

*Claim.* Let $H$ be a proper interval graph, and $P = p_1, \ldots, p_\ell$ be an induced path in $H$. Let $u \notin \{p_1, \ldots, p_\ell\}$ be some vertex of $H$ and let $N_P(u)$ be the set of its neighbours in $P$. Then, the vertices of $|N_P(u)|$ occur consecutively on the path $P$, and furthermore, $|N_P(u)| \leq 4$.

*Proof.* The first statement follows from the fact that $H$ has no induced cycle of length more than three and the second statement from the fact that $H$ contains no claw. □

Let $(G, k)$ be a YES instance and let $X \subseteq V(G)$ be a vertex set such that $|X| \leq k$ and $G \setminus X$ is a proper interval graph. Then clearly $(G \setminus v, k)$ is a YES instance of PIVD as $|X \setminus \{v\}| \leq k$ and $G \setminus (\{v\} \cup X)$ is a proper interval graph.

For the opposite direction, let $(G \setminus v, k)$ be a YES instance for PIVD and let $X$ be a vertex set such that $|X| \leq k$ and $G \setminus (\{v\} \cup X)$ is a proper interval graph. Towards a contradiction, let us assume that $G \setminus X$ is not a proper interval graph. Thus $G \setminus X$ contains one of the forbidden induced subgraphs for proper interval graphs. We first show that this can not contain forbidden induced subgraphs of size at most 8. Let $Y$ be the subset of $X$ such that it is a minimal hitting set for nets, tents, claws and induced cycles $C_\ell$, $4 \leq \ell \leq 8$, contained in $G[T]$. By the definition of $T$ and the fact that $v \notin T$ we know that $Y$ is also a minimal hitting set for nets, tents, claws and induced cycles $C_\ell$, $4 \leq \ell \leq 8$, contained in $G$. Thus, the only possible candidate for the forbidden subgraph in $G \setminus X$ is an induced cycle $C_\ell$, where $\ell \geq 9$. Now since $G \setminus (X \cup \{v\})$ is a proper interval graph, the vertex $v$ is part of the cycle $C_\ell = \{v, w_1, w_2, \ldots, w_\ell\}$. Furthermore, $w_1$ and $w_\ell$ are the neighbors of $v$ on $C_\ell$.

Next we show that using $C_\ell$ we can construct a forbidden induced cycle in $G \setminus (\{v\} \cup X)$, contradicting that $G \setminus (\{v\} \cup X)$ is a proper interval graph. Towards this we proceed as follows. For vertex sets $V^F$ and $V^L$ (the first and the last $k+1$ vertices of $\sigma(K)$), we pick up vertices $v^F \in V^F \setminus X$ and vertex $v^L \in V^L \setminus X$. Because $|X| \leq k$, such vertices always exist.

*Claim.* Vertices $w_1, w_\ell \in T \cup N[v^F] \cup N[v^L]$.

*Proof.* Let $w_a \in (T \cup K)$, $a \in \{1, \ell\}$, then because $K \subseteq N[v^F]$, we are done. Otherwise $w_a$ is a vertex of the proper interval graph $G_T \setminus K$. Then $w_a$ occurs either before or after the vertices of $K$ in $\sigma$. If $w_a$ occurs before then $w_a < v^F < v$ on $\sigma$. Now since $w_a$ has an edge to $v$ and $\sigma$ is a proper interval ordering of $G_T$, we have that $w_a v^F$ is an edge and hence $w_a \in N[v^F]$. The case where $w_a$ occurs after $K$ is symmetric. In this case we could show that $w_a \in N[v^L]$. □

Now with $w_a, a \in \{1, \ell\}$, we associate a partner vertex $p(w_a)$. If $w_a \in T$, then $Z \cap N(w_a)$ contains at least $k+1$ vertices as $v \in K \cap N(w_a)$ is not in $Z$. Thus there exists $z^a \in (Z \cap N(w_a)) \setminus X$. In this case we define $p(w_a)$ to be $z^a$. If $w_a \notin T$ then by Claim 4.2 we know that either $v^F$ or $v^L$ is a neighbor to $w_a$. If $v^F$ is neighbor to $w_a$ then we define $p(w_a) = v^F$, else $p(w) = v^L$. Observe that $p(w_a) \in K \setminus \{v\}$ for $a \in \{1, \ell\}$.

Now consider the closed walk $W = \{p(w_1), w_1, \ldots, w_\ell, p(w_\ell)\}$ in $G \setminus (\{v\} \cup X)$. First of all $W$ is a closed walk because $p(w_1)$ and $p(w_\ell)$ are adjacent. In fact, we would like to show that $W$ is a simple cycle in $G \setminus (\{v\} \cup X)$ (not necessarily an induced cycle). Towards this we first show that $p(w_a) \notin \{w_1, \ldots, w_\ell\}$. Suppose $p(w_1) \in \{w_1, \ldots, w_\ell\}$, then it must be $w_2$ as the only neighbors of $w_1$ on $C_\ell$ are $v$ and $w_2$. However, $v$ and $p(w_1)$ are part of the same clique $K$. This implies that $v$ has $w_1, w_2$ and $w_\ell$ as its neighbors on $C_\ell$, contradicting to the fact that $C_\ell$ is an induced cycle of length at least 9 in $G$. Similarly, we can also show that $p(w_\ell) \notin \{w_1, \ldots, w_\ell\}$. Now, the only reason $W$ may not be a simple cycle is that $p(w_1) = p(w_\ell)$. However, in that case $W = \{p(w_1), w_1, \ldots, w_\ell\}$ is a simple cycle in $G \setminus (\{v\} \cup X)$.

Notice that $G[\{w_1, w_2, \ldots, w_\ell\}]$ is an induced path, where $\ell \geq 8$. Let $i$ be the largest integer such that $w_i \in N(p(w_1))$ and let $j$ be the smallest integer such that $w_j \in N(p(w_\ell))$. By Claim 4.2 and the conditions that $G[\{w_1, w_2, \ldots, w_\ell\}]$ is an induced path, $w_1 \in N(p(w_1))$, and $w_\ell \in N(p(w_\ell))$, we get that $i \leq 4, j \geq \ell - 3$. As $\ell \geq 8$ this implies that $i < j$, and hence $G[\{v_1, w_i, \ldots, w_j, v_\ell\}]$ is an induced cycle of length at least four in $G \setminus (\{v\} \cup X)$, which is a contradiction. Therefore, $G \setminus X$ is a proper interval graph. □

### 4.3 Shrinking $G_T$

Let $(G, k)$ be a YES instance of PIVD, and let $T$ be a vertex subset of $G$ of size at most $\delta(k)$ such that $G_T = G \setminus T$ is a proper interval graph with the maximum clique size at most $\epsilon(k) = (k+1)(\delta(k) + 2)$. The following lemma argues that if $G_T$ has sufficiently long clique path, then a part of this path can be shrunk without changing the solution.

**Lemma 5.** *Let us assume that every claw in $G$ contains at least two vertices from $T$ and that there is a connected component of $G_T$ with at least $\zeta(k) = \Big(\delta(k)(8\epsilon(k) + 2) + 1\Big)\Big(2[\epsilon(k)]^2 + 32\epsilon(k) + 3\Big)$ maximal cliques. Then there is a polynomial time algorithm transforming $G$ into a graph $G'$ such that*

- *$(G, k)$ is a YES instance if and only if $(G', k)$ is a YES instance;*
- *$|V(G')| < |V(G)|$.*

*Proof.* Let $I$ be a connected component of $G_T$ with $\zeta(k)$ maximal cliques and let $C_1, C_2, \ldots, C_p$, $p \geq \zeta(k)$, be a clique path of $I$. We first show that every vertex $v$ of $I$ belongs to at most $2\epsilon(k)$ maximal cliques of $I$. We know that every clique of $I$ has size at most $\epsilon(k)$ and by the property 2 of Proposition 3 we have that the neighborhood of any vertex $v$ can be covered with at most 2 cliques. Thus, $|N_I(v)| \leq 2\epsilon(k)$. Now, by the last property of Proposition 3, we have that if $j$ is the least integer such that $v \in C_j$, then $v \notin C_{j+\ell+1}$ for $\ell \geq |N_I(v)|$. This proves our claim. For every vertex $v \in T$, we mark all maximal cliques of $I$ containing at least one neighbour of $v$. Let $m(v)$ be the set of maximal cliques marked for vertex $v$. We claim that

$$\forall v \in T, |m(v)| \leq 8\epsilon(k) + 2. \quad (1)$$

Every vertex of $I$ is in at most $2\epsilon(k)$ maximal cliques, and thus in every set of $4\epsilon(k)+1$ maximal cliques, no vertex of the first clique (in the ordering of the clique path) can be adjacent to a vertex of the last clique. Thus in every set of $8\epsilon(k)+2$ maximal cliques, it is always possible to select three cliques such that no vertex of one clique is adjacent to a vertex of another. Thus, if $m(v)$ consisted of at least $8\epsilon(k)+2$ maximal cliques, vertex $v$ would have at least three neighbours in $I$ which are pairwise non-adjacent. In other words, these three vertices together with $v$, form a claw. This claw contains exactly one vertex from $T$, contradicting the assumption that every claw in $G$ has at least two vertices from $T$. This proves (1).

By (1), the total number of marked maximal cliques in $I$ is at most $|T|(8\epsilon(k)+2) = \delta(k)(8\epsilon(k)+2)$. By the pigeonhole principle, the clique path $C_1, C_2, \ldots, C_p$ contains at least

$$\ell = \frac{\zeta(k)}{\delta(k)(8\epsilon(k)+2)+1} = 2[\epsilon(k)]^2 + 32\epsilon(k) + 3$$

consecutive unmarked maximal cliques, i.e. cliques containing no vertices adjacent to vertices of $T$. Let $C_i, C_{i+1}, \ldots, C_{i+\ell-1}$ be a set of unmarked consecutive cliques. Let $q = 16\epsilon(k)+1$. Then for every $v \in C_i$ and $u \in C_{i+q}$, the distance between $v$ and $u$ in $G$ is at least 9. This is because every shortest path from $v$ to $u$ should contain at least one vertex either from each of the cliques from $C_i, C_{i+1}, \ldots, C_{i+q}$ or from each of the cliques $C_{i+q}, C_{i+q+1}, \ldots, C_{i+\ell-1}$. In both cases, every shortest path goes through at least $q = 16\epsilon(k)+1$ maximal cliques, and then by Proposition 3, the length of such path is at least 9. By similar arguments, for every $v \in C_{i+\ell-1-q}$ and $u \in C_{i+\ell-1}$, the distance between $v$ and $u$ in $G$ is also at least 9.

Clique $C_{i+q}$ is maximal, and thus it contains a vertex $x$ which does not belong to $C_{i+q+1}$. Similarly, let $y \in C_{i+\ell-1-q} \setminus C_{i+\ell-1-q-1}$. We compute the minimum size $s$ of an $x, y$-separator in $I$. Let us note, that in proper interval graphs such a separator is an intersection of two consecutive cliques in the clique path, and can be found in polynomial time. We construct a new graph $G'$ from $G$ as follows. In this new graph $G'$, the connected component $I$ of proper interval graph $G_T$ is replaced by a smaller proper interval graph $I'$. The proper interval graph $I'$ is formed by cliques $C_1, C_2, \ldots, C_{i+q}$ and $C_{i+\ell-1-q}, C_{i+\ell-q}, \ldots, C_p$, and a new clique $C$ of size $s$. We make all vertices of $C$ adjacent to all vertices of $C_{i+q}$ and of $C_{i+\ell-1-q}$. Let us note that because there are at least $2[\epsilon(k)]^2 + 1$ maximal cliques in clique path between $C_{i+q}$ and $C_{i+\ell-1-q}$, there are at least $\epsilon(k)+1$ vertices belonging only to these cliques. Thus $I'$ has less vertices than $I$.

It is easy to check that $I'$ is a proper interval graph. The construction of the graph $I'$, and hence the graph $G'$, can be done in polynomial time. Indeed, marking maximal cliques can be performed in polynomial time, and then computing the size of a minimal separator in $I$ can be also performed in polynomial time.

What remains is to argue that $(G, k)$ is a YES instance if and only if $(G', k)$ is a YES instance. Let $R = V(I) \setminus (V(I') \cup C)$ be the vertices of $G$ removed during construction of graph $G'$. Vertices of $R$ cannot be contained in any of the forbidden small induced graphs: claw, net, tent, and cycles $C_\ell$, $\ell \leq 8$. The reason to that is that the distance from any vertex $v$ of $R$ to any vertex of $T$ is at least 9, and thus if such a small induced graph contains $v$, it should be a subgraph of the proper interval graph $I$, which is a contradiction. Thus every set which hits small forbidden induced subgraphs in $G$ also hits them in $G'$ and vice versa. So in further arguments we concentrate only on induced cycles (holes) of length at least 9.

Let $X$ be a set of size at most $k$ such that $G \setminus X$ is a proper interval graph. We assume that the set $X$ is a minimal set with such properties. If $X$ intersects $R$, this is because $X$ hits some cycles passing in $G$ through $R$. We claim that in this case, $X$ should contain at least $s$ vertices from $R$. Let $v \in X \cap R$. By minimality of $X$, there should be a witness hole $Q$ of length at least 9 and such that $Q \cap X = \{v\}$. Let $u$ and $w$ be the neighbours of $v$ in $Q$. Then the set of vertices $N_G(u) \cap N_G(w)$ is a subset of $X$. Indeed, if there was a vertex $v' \in (N_G(u) \cap N_G(w)) \setminus X$, then the cycle obtained from $Q$ by replacing $v$ with $v'$ is a hole of length at least 9 avoiding $X$. On the other hand, by making use of Proposition 3, it easy to show that the set $N_G(u) \cap N_G(w)$ separates $x$ and $y$, and thus $|N_G(u) \cap N_G(w)| \geq s$. Therefore either $|X \cap R| \geq s$, or $X \cap R = \emptyset$.

If $|X \cap R| \geq s$, then $X' = (X \setminus R) \cup C$ is of size at most $|X|$. We claim that $G' \setminus X'$ is a proper interval graph. Indeed, every induced subgraph of $G'$ which is forbidden for proper interval graphs either intersects $C$, or should be also a subgraph of $G \setminus R$. But in both cases, this subgraph is hit by $X'$.

Suppose that $R \cap X = \emptyset$. We know that $G' \setminus X'$ cannot contain a claw, net, tent, and induced cycles $C_\ell$, $\ell \leq 8$, because each of these subgraphs is entirely in $G \setminus R$, and thus is hit by $X'$. If $G' \setminus X'$ contain an induced cycle $Q$ of length at least 9, then this cycle cannot be entirely in $G \setminus R$ and thus should touch $C$. Because $Q$ is an induced cycle, it should contain a path passing through a vertex $a \in C_{i+q}$, then continuing trough a vertex from $C$, and through a vertex $b \in C_{i+\ell-1-q}$. In $G$ vertices $a$ and $b$ are also connected by a path whose vertices use only vertices of $R$, and thus avoiding $X$. Replacing in $Q$ the $a,b$-path passing through $C$ by an $a,b$-path passing through $R$, we obtain an induced cycle of length at least 9 in $G$ avoiding $X$. But this is a contradiction, and we conclude that $G' \setminus X'$ is a proper interval graph. We have shown that if $(G, k)$ is a YES instance then $(G', k)$ is a YES instance.

Let $X'$ be a minimal proper interval vertex deletion set of graph $G'$. If $X'$ intersects $C$, let $v \in X' \cap C$. Because $X'$ is minimal, we can select a witness induced cycle $Q$ of length at least 9 such that $v$ is the only vertex of $X'$ in $Q$. Let $u \in C_{i+q}$, $w \in C_{i+\ell-1-q}$ be the neighbours of $v$ in $Q$. Then every vertex from the set $N(u) \cap N(w)$ should be in $X'$ too because otherwise it would be possible to modify $Q$ into an induced cycle $Q$ of length at least 9 avoiding $X$. By the way we constructed graph $G'$, we have that $C = N(u) \cap N(w)$, and thus $C \subseteq X$. Let $C'$ be a minimum $x,y$-separator in $G$. The size of $C'$ is $s$, and the set $X = (X' \setminus C) \cup C'$ is of size $|X'|$. Every forbidden (for proper interval graph) subgraph in $G$ avoiding $X' \setminus C$ should contain a path connecting a vertex from $C_{i+q}$ to a vertex from $C_{i+\ell-1-q}$, and thus is hit by $C'$. Thus $G \setminus X$ is a proper interval graph. If $X'$ does not intersect clique $C$, then $G \setminus X'$ is a proper interval graph. Indeed, every induced cycle of length at least 9 in $G$ containing a path $P$ connecting a vertex from $C_{i+q}$ to a vertex from $C_{i+\ell-1-q}$, can be transformed to a cycle of length at least 9 in $G'$ by replacing $P$ with a path of length 2 passing through $C$. This implies that $X'$ hits every forbidden subgraph in $G$ too. We have shown that if $(G', k)$ is a YES instance then $(G, k)$ is a YES instance, which concludes the proof of the lemma. □

### 4.4 Putting all together: final kernel analysis

We need some auxiliary reduction rules to give the kernel for PIVD. Let $\mathcal{F}$ be the family consisting of all nets, tents, claws and induced cycles $C_\ell$ for $\ell \in \{4, 5, \ldots, 8\}$ of the input graph $G$.

**Lemma 6.** *Let $(G, k)$ be an instance to PIVD and $T$ be as defined before. Let $X$ be a subset of $T$ such that for every $x \in X$ we have a set $S_x \in \mathcal{F}$ such that $S_x \setminus \{x\} \subseteq (V(G) \setminus T)$. If $|X| > k$ then we conclude that $G$ can not be transformed into proper interval graph by deleting at most $k$ vertices. Else, $(G, k)$ is a YES instance if and only if $(G[V \setminus X], k - |X|)$ is a YES instance.*

*Proof.* We first argue that $X$ is a subset of every minimal hitting set $S'$ of size at most $k$ for $\mathcal{F}$. By the property of the set $T$ we have that $S' \subseteq T$. This implies that any forbidden set that is contained in $\mathcal{F}$ such that all but one of its vertex is in $T$ must be contained in every minimal hitting set of size at most $k$. This shows that $X \subseteq S'$.

Suppose $(G,k)$ is YES instance. Then there exists a set $P$ of size at most $k$ such that $G \setminus P$ is a proper interval graph. Hence this is also a hitting set for $\mathcal{F}$. Let $P'$ be a subset of $P$ such that $P'$ is a minimal hitting set for $\mathcal{F}$. By the property of the set $T$ we have that $P' \subseteq T$. By the arguments in first paragraph we have that $X \subseteq P'$. In fact, $X$ is subset of every subset of size at most $k$ such that its deletion makes the graph proper interval. Hence $(G,k)$ is a YES instance if and only if $(G \setminus X, k - |X|)$ is a YES instance. This also implies that if $|X| > k$ then $(G,k)$ is a NO instance and hence in this case we conclude that $G$ can not be made into proper interval graph by deleting at most $k$ vertices. This completes the proof. □

Now we are ready to state the main result of this paper.

**Theorem 1.** PIVD *admits a polynomial kernel.*

Before proceedings with the proof of the theorem, let us remind the definitions of all functions used so far.

---

Size of $T$: $\leq \delta(k) = 8 \cdot 8!(k+1)^8 + k$
Maximum clique size in $G_T$: $\leq \epsilon(k) = (k+1)(\delta(k) + 2)$
\# of maximal cliques in $G_T$: $\leq \zeta(k) = \Big(\delta(k)(8\epsilon(k) + 2) + 1\Big)\Big(2[\epsilon(k)]^2 + 32\epsilon(k) + 3\Big)$

---

*Proof.* Let $(G,k)$ be an instance to PIVD. We first show that if $G$ is not connected then we can reduce it to the connected case. If there is a connected component $\mathcal{C}$ of $G$ such that $\mathcal{C}$ is a proper interval graph then we delete this component. Clearly, $(G,k)$ is a YES instance if and only if $(G \setminus \mathcal{C}, k)$ is a YES instance. We repeat this process until every connected component of $G$ is not a proper interval graph. At this stage if the number of connected components is at least $k+1$, then we conclude that $G$ can not be made into a proper interval graph by deleting at most $k$ vertices. Thus, we assume that $G$ has at most $k$ connected components. Now we show how to obtain a kernel for the case when $G$ is connected, and for the disconnected case we just run this algorithm on each connected component. This only increases the kernel size by a factor of $k$. From now onwards we assume that $G$ is connected.

Now we apply Lemma 3 on $G$ and in polynomial time either find a non-empty set $T \subseteq V(G)$ such that

1. $G \setminus T$ is a proper interval graph;
2. $Y \subseteq V(G)$ of size at most $k$ is a minimal hitting set for nets, tents, claws and induced cycles $C_\ell$ for $\ell \in \{4, 5, \ldots, 8\}$ contained in $G$ if and only if it is a minimal hitting set for nets, tents, claws and induced cycles $C_\ell$ for $\ell \in \{4, 5, \ldots, 8\}$ contained in $G[T]$; and
3. $|T| \leq \delta(k)$,

or conclude that $G$ can not be made into a proper interval graph by deleting at most $k$ vertices. If Lemma 3 concludes that $G$ can not be transformed into a proper interval graph by deleting at most $k$ vertices, then the kernelization algorithm returns the same.

If the size of a maximum clique in $G_T$ is more than $\epsilon(k)$, then we apply Lemma 4 and obtain a vertex $v \in V(G_T)$ such that $(G,k)$ is a YES instance if and only if $(G \setminus v, k)$ is a YES instance. We apply Lemma 4 repeatedly until the size of a maximum clique in $G_T$ is at most $\epsilon(k)$. So, from now onwards we assume that the size of a maximum clique in $G_T$ is at most $\epsilon(k)$.

Now we apply Lemma 6 on $(G, k)$. If Lemma 6 concludes that $G$ cannot be made into a proper interval graph by deleting at most $k$ vertices, then $(G, k)$ is a NO instance and the kernelization algorithm returns a trivial NO instance. Otherwise, we find a set $X \subseteq T$ such that $(G, k)$ is a YES instance if and only if $(G \setminus X, k - |X|)$ is a YES instance. If $|X| \geq 1$ then $(G \setminus X, k - |X|)$ is a smaller instance and we start all over again with this as new instance to PIVD.

If we cannot apply Lemma 6 anymore, then every claw in $G$ contains at least two vertices from $T$. Thus if the number of maximal cliques in a connected component of $G_T$ is more than $\zeta(k)$, we can apply Lemma 5 on $(G, k)$ and obtain an equivalent instance $(G', k)$ such that $|V(G')| < |V(G)|$ and then we start all over again with instance $(G', k)$.

Finally, we are in the case, where $G_T$ is a proper interval graph and none of conditions of Lemmata 4, 5 and 6 can be applied. This implies that the number of maximal cliques in each connected component of $G_T$ is at most $\zeta(k)$ and the size of each maximal clique is at most $\epsilon(k)$. Thus we have that every connected component of $G_T$ has at most $\zeta(k)\epsilon(k)$ vertices. Since $G$ is connected, we have that every connected component of $G_T$ has some neighbour in $T$. However because Lemma 6 cannot be applied, we have that every vertex in $T$ has neighbours in at most 2 connected components. The last assertion follows because of the following reason. If a vertex $v$ in $T$ has neighbours in at least 3 connected components of $G_T$ then $v$ together with a neighbour from each of the components of $G_T$ forms a claw in $G$, with all the vertices except $v$ in $G_T$, which would imply that Lemma 6 is applicable. This implies that the total number of connected components in $G_T$ is at most $2\delta(k)$. Thus the total number of vertices in $G$ is at most $2\delta(k)\zeta(k)\epsilon(k)$.

Recall that $G$ may not be connected. However, we argued that $G$ can have at most $k$ connected components and we apply the kernelization procedure on each connected component. If the kernelization procedure returns that some particular component can not be made into a proper interval graph by deleting at most $k$ vertices, then we return the same for $G$. Else, the total number of vertices in the reduced instance is at most $2k \cdot \delta(k)\zeta(k)\epsilon(k)$, which is a polynomial.

Observe that the above procedure runs in polynomial time, as with every step of the algorithm, the number of vertices in the input graph reduces. This together with the fact that Lemmata 4, 5 and 6 run in polynomial time, we have that the whole kernelization algorithm runs in polynomial time. This concludes the proof. □

## 5 Conclusion and discussions

In this paper we proved that PIVD admits a polynomial kernel. While resolving the complexity of the problem from kernelization perspective, we have to admit that in the current form our result is purely of theoretical importance. This is due to the large number $2k \cdot \delta(k)\zeta(k)\epsilon(k) \in O(k^{53})$ of vertices in our kernel. It is possible to improve the sizes of our kernel slightly by the cost of tedious case analysis but the challenging open question is if a kernel of "reasonably" polynomial size, say $k^{10}$ is possible. It seams that for this type of a kernel we need completely different techniques. On the other hand, is it possible to prove that PIVD does not have a kernel of size $k^7$?

Another interesting open question is if $p$-CHORDAL GRAPH VERTEX DELETION admits a polynomial kernel. The problem is known to be FPT by the result of Marx [19]. And finally, what about $p$-INTERVAL GRAPH VERTEX DELETION? We even do not know if the problem is FPT.